\documentclass[prd,preprint,showpacs,preprintnumbers,amsmath,amssymb,eqsecnum]{revtex4}

\usepackage{graphicx}
\usepackage{dcolumn}
\usepackage{bm}
\usepackage{subfigure}
\usepackage{color}

\begin{document}

\preprint{RUP-10-2}
\preprint{OCU-PHYS 338}
\preprint{AP-GR 82}

\title{
Collision of an innermost stable circular orbit 
particle around a Kerr black hole
}
\author{$^{1}$Tomohiro Harada}
\email{harada@rikkyo.ac.jp}
\author{$^{2}$Masashi Kimura}
\email{mkimura@sci.osaka-cu.ac.jp}
\affiliation{%
$^{1}$Department of Physics, Rikkyo University, Toshima, Tokyo 175-8501, Japan\\
$^{2}$Department of Mathematics and Physics, Graduate School of Science, Osaka City University, Osaka 558-8585, Japan
}%
\date{\today}

\begin{abstract}
We derive a general formula for the center-of-mass (CM) energy  
for the near-horizon collision of two particles
of the same rest mass on the equatorial plane around
a Kerr black hole. We then apply this formula
to a particle which plunges from the innermost stable circular orbit (ISCO) 
and collides with another particle
near the horizon.
It is found that 
the maximum value of the CM energy $E_{\rm cm}$ 
is given by $E_{\rm cm}/(2m_{0})
\simeq 1.40/\sqrt[4]{1-a_{*}^{2}}$
for a nearly maximally rotating black hole, 
where  $m_{0}$ is the rest mass of each particle and  
$a_{*}$ is the nondimensional Kerr parameter.
This coincides 
with the known upper bound for a particle which begins at rest 
at infinity within a factor of 2.
Moreover, we also consider the collision of a particle 
orbiting the ISCO with another particle on the ISCO and find that the 
maximum CM energy is then given by 
$E_{\rm cm}/(2m_{0})\simeq 1.77/\sqrt[6]{1-a_{*}^{2}}$.
In view of the astrophysical significance of the ISCO, 
this result implies that 
particles can collide around 
a rotating black hole with an arbitrarily high CM energy
without any artificial fine-tuning in an astrophysical context 
if we can take the maximal limit of the black hole spin 
or $a_{*}\to 1$.
On the other hand, even if we take Thorne's bound on the 
spin parameter into account, highly or moderately 
relativistic collisions
are expected to occur quite naturally, for  
$E_{\rm cm}/(2m_{0})$ takes 6.95 (maximum) and 3.86 (generic)
near the horizon and 4.11 (maximum) and 2.43 (generic) on the ISCO
for $a_{*}=0.998$.
This implies that high-velocity collisions of compact objects are naturally expected around a rapidly rotating supermassive black hole.
Implications to accretion flows onto a rapidly rotating black hole 
are also discussed.
\end{abstract}

\maketitle


\section{Introduction}
Recently, Banados, Silk and West~\cite{BSW2009} 
showed that if
two particles which begin at rest at infinity 
collide near the horizon of a maximally rotating Kerr 
black hole~\cite{Kerr1963} and if the angular momentum of either particle
is fine-tuned, the center-of-mass (CM) 
energy $E_{\rm cm}$ of the two particles
can be arbitrarily high 
and hence the maximally rotating black hole might be regarded as 
a Planck-energy-scale collider.
We here call this the Banados-Silk-West (BSW) effect.
This scenario was subsequently 
criticized~\cite{Berti_etal2009,Jacobson_Sotiriou2010}
from several points, such as 
astrophysical bounds on the black hole spin parameter,
the effects of gravitational waves, the self-gravity of the particles
and the long proper time needed for such a collision.
In the near-maximal rotation, the CM energy 
of two particles of mass $m_{0}$ is bounded by 
$E_{\rm cm}/(2m_{0})\sim 2.41/\sqrt[4]{1-a_{*}^{2}}$, 
where $a_{*}$ is the nondimensional 
Kerr parameter~\cite{Berti_etal2009,Jacobson_Sotiriou2010}.
On the other hand, 
Grib and Pavlov~\cite{Grib_Pavlov2010_Kerr}  
proposed a slightly different mechanism 
that $E_{\rm cm}$ can be arbitrarily high
even in the non-maximal rotation if the particle has experienced
multiple scattering and fine-tuned its angular momentum before 
the relevant collision.
The BSW effect is also analyzed in 
the Kerr-Newman family of black holes~\cite{WLGF2010}, 
general stationary and axisymmetric black holes~\cite{Zaslavskii2010_rotating},
and 
static charged black holes~\cite{Zaslavskii2010_charged}.

In the BSW effect, to obtain such an arbitrarily high $E_{\rm cm}$, 
the angular momentum 
of either particle must be fine-tuned. However, there is a natural 
mechanism in astrophysics to tune the particle's energy and 
angular momentum.
This is the innermost stable circular orbit (ISCO).
The ISCO around a Kerr black hole is studied in detail 
by Bardeen, Press and Teukolsky~\cite{BPT1972}.

In the geometrically thin and optically thick accretion disk 
model~\cite{Shakura_Sunyaev1973,Page_Thorne1974}, 
which is known as 
the standard accretion disk model,
a rotating fluid or plasma gradually takes
a circular orbit which is closer to the black hole
as the fluid transfers its angular momentum outwards and releases its energy 
by electromagnetic radiation in the time scale of viscosity,
which is much longer than the black hole dynamical time scale.
This electromagnetic emission can be observed 
by radio interferometers and X-ray observation satellites. 
Once the fluid reaches the inner edge of the accretion disk,
which is given by the ISCO, 
it begins to plunge into the black hole in the dynamical 
time scale~\footnote{
The fluid will not take a 
quasicircular orbit around the ISCO in the radiatively 
inefficient accretion flows,
including the advection-dominated accretion flows, as shown 
in e.g.~\cite{Takahashi2007}. See also~\cite{Kato_etal2008} for 
the accretion disk models.}.
In the plunging phase,
its energy and angular momentum are approximately 
conserved.
It should be noted that due to the accretion of radiation emitted 
from the disk, there is an astrophysical upper bound on the 
nondimensional Kerr parameter, what we call  
Thorne's bound, $|a_{*}|\lesssim 0.998$~\cite{Thorne1974}.

Another important example where the ISCO plays a crucial role is 
inspirals of stellar-mass compact objects 
into supermassive black holes, which 
are called extreme mass-ratio inspirals.
Extreme mass-ratio inspirals are interesting sources of 
gravitational waves for the Laser Interferometric 
Space Antenna~\cite{LISA2000}.
If instead the central mass is an intermediate-mass black hole,
these are interesting sources for 
the DECi-hertz Interferometer Gravitational wave Observatory~\cite{Seto_etal2001,Kawamura_etal2006}
and the Big Bang Observatory~\cite{Phinney_etal2004,Crowder_Cornish2005}. 
Also in this case, an inspiralling compact object 
gradually takes a circular orbit which is closer to the black hole
as the object transfers its angular momentum outwards and releases its energy 
by gravitational waves in the time scale of gravitational radiation,
which is much longer than the dynamical time scale.
Once the compact object reaches the ISCO,
it begins to plunge into the black hole in the dynamical time scale.
In the plunging phase, 
its energy and angular momentum are approximately conserved.

In this paper, we derive a general formula for the CM energy
for the near-horizon collision of two particles on the equatorial plane
around a Kerr black hole, which is valid in both the maximal and 
non-maximal rotation cases. 
Then, we apply this formula to  
the near-horizon collision of two particles, either 
of which is plunging from the ISCO. We find that the BSW 
effect occurs in the near-maximal rotation and that the 
maximum value for the CM energy of the ISCO particles 
is the same 
as the upper bound for the particles initially at rest at infinity within 
a factor of 2.
This implies that the BSW effect is not an artificial 
but physically realistic astrophysical phenomenon.
We also consider the collision of a particle orbiting the ISCO
with another generic particle on the ISCO and find that the associated CM
energy can also be arbitrarily high in exactly the same sense as BSW's,
although the dependence on the black hole spin parameter is 
quite different.
We neglect the effects of gravitational waves
and the self-gravity of the particles.

This paper is organized as follows. 
In Sec.~II we briefly review particle orbits and the CM energy 
for the collision of two particles in the Kerr spacetime.
In Sec.~III, we discuss particle orbits near the horizon and 
derive a general formula  
for the CM energy 
for the near-horizon particle collision.
In Sec.~IV, we apply this formula to a particle which plunges from the 
ISCO and obtain the CM energy for different collisions.
In Sec.~V, we investigate the collision of a particle orbiting the 
ISCO with another particle on the ISCO.
Section VI is devoted to conclusion and discussion.
We use the units in which $c=G=1$ and 
the abstract index notation of Wald~\cite{Wald1984}.
\section{CM energy for particle collision in the Kerr spacetime}
In this section, we briefly review particle orbits and the CM energy 
for the two-particle collision on the equatorial 
plane of the Kerr spacetime in the general situation, 
following~\cite{BSW2009,Jacobson_Sotiriou2010,Grib_Pavlov2010_Kerr}. 
We use a similar notation to that of 
Grib and Pavlov~\cite{Grib_Pavlov2010_Kerr}.

\subsection{Particle orbits in the Kerr spacetime}
\label{subsec:particle_orbits}
The line element in the Kerr spacetime
in the Boyer-Lindquist coordinates is given by~\cite{Kerr1963,Wald1984,Poisson2004}  
\begin{eqnarray*}
ds^{2}&=&-\left(1-\frac{2Mr}{\rho^{2}}\right)dt^{2}
-\frac{4Mar\sin^{2}\theta}{\rho^{2}}d\phi dt
+\frac{\rho^{2}}{\Delta}dr^{2}+\rho^{2}d\theta^{2} \nonumber \\
&&+\left(r^{2}+a^{2}+\frac{2Mra^{2}\sin^{2}\theta}{\rho^{2}}\right)
\sin^{2}\theta d\phi^{2} ,
\label{eq:Kerr_metric}
\end{eqnarray*}
where $a$ and $M$ are, respectively, the spin and mass parameters, 
$\rho^{2}=r^{2}+a^{2}\cos^{2}\theta$ and $\Delta=r^{2}-2Mr+a^{2}$.
We assume $a\ge 0$ without loss of generality. 
If $a^{2}\le M^{2}$, 
$\Delta$ vanishes at $r=r_{\pm}=M\pm\sqrt{M^{2}-a^{2}}$, where 
$r=r_{+}$ and $r=r_{-}$ correspond to an event horizon and 
a Cauchy horizon, respectively. Here, we denote $r_{+}=r_{H}$
and $r_{-}=r_{C}$. In this coordinate system, the time 
translational and the axial Killing vectors 
are respectively given by
\begin{equation*}
\xi^{a}=\left(\frac{\partial}{\partial t}\right)^{a},\quad 
\psi^{a}=\left(\frac{\partial}{\partial \phi}\right)^{a}.
\end{equation*}
The surface gravity of the Kerr black hole is given by
\begin{equation*}
\kappa=\frac{\sqrt{M^{2}-a^{2}}}{r_{H}^{2}+a^{2}}.
\end{equation*}
Thus, the black hole has a vanishing surface gravity and hence is 
extremal for the maximal rotation $a^{2}=M^{2}$, while 
it is subextremal for the non-maximal rotation $a^{2}<M^{2}$. 
The angular velocity of the 
horizon is given by 
\begin{equation*}
\Omega_{H}=\frac{a}{r_{H}^{2}+a^{2}}=\frac{a}{2M(M+\sqrt{M^{2}-a^{2}})}.
\end{equation*}
The Killing vector $\chi^{a}=\xi^{a}+\Omega_{H}\psi^{a}$ is a null generator of the event horizon.

Here we concentrate on particles which move on the 
equatorial plane $\theta=\pi/2$.
Then, the four velocity $u^{a}=\dot{x}^{a}$ of the particle has a vanishing $\theta$ component, i.e., $u^{\theta}=0$, 
where the dot denotes the differentiation with respect to the 
affine parameter of the geodesic.
From Eq.~(\ref{eq:Kerr_metric}), the line element on the equatorial plane in the Kerr spacetime 
is given by 
\begin{equation}
ds^{2}=-\left(1-\frac{2M}{r}\right)dt^{2}
-\frac{4aM}{r}dtd\phi
+\frac{r^{2}}{\Delta}dr^{2}
+\left(r^{2}+a^{2}+\frac{2Ma^{2}}{r}\right)d\phi^{2}.
\label{eq:equatorial_BLmetric}
\end{equation}
Associated with the Killing vectors $\xi^{a}$
and $\psi^{a}$, we have 
the following conserved quantities along a geodesic on the equatorial plane:
\begin{eqnarray}
e&=&-g_{ab}\xi^{a}u^{b}=-u_{t}=-(g_{tt}u^{t}+g_{t\phi}u^{\phi}),
\label{eq:e}\\
L&=&g_{ab}\psi^{a}u^{b}=u_{\phi}=g_{\phi t}u^{t}+g_{\phi\phi}u^{\phi},
\label{eq:L}
\end{eqnarray}
where $e$ and $L$ correspond the specific energy and angular momentum, respectively.
Solving the above for $u^{t}$ and $u^{\phi}$, we have
\begin{eqnarray}
\dot{t}&=&\frac{1}{\Delta}\left[\left(r^{2}+a^{2}+\frac{2Ma^{2}}{r}\right)e
-\frac{2Ma}{r}L\right], 
\label{eq:tdot}\\
\dot{\phi}&=&\frac{1}{\Delta}\left[\left(1-\frac{2M}{r}\right)L 
+\frac{2Ma}{r}e\right].
\label{eq:phidot}
\end{eqnarray}
To have $\dot{t}\ge 0$, the condition
\begin{equation*}
\left(r^{2}+a^{2}+\frac{2Ma^{2}}{r}\right)e
-\frac{2Ma}{r}L\ge 0
\end{equation*}
must be satisfied outside the event horizon. In the limit to 
the horizon $r\to r_{H}$ from outside, this
condition reduces to 
\begin{equation*}
l\le l_{H}=\frac{2(1+\sqrt{1-a_{*}^{2}})}{a_{*}}e=\frac{e}{M\Omega_{H}},
\end{equation*}
or 
\begin{equation*}
e\ge \Omega_{H}L,
\end{equation*}
where we put $a_{*}=a/M$ and $l=L/M$. In terms of $a_{*}$, 
$0\le a_{*}<1$ for the subextremal case, while $a_{*}=1$ for the 
extremal case.

Substituting Eqs.~(\ref{eq:tdot}) and (\ref{eq:phidot}) 
into the normalization condition $u^{a}u_{a}=-1$ with $\theta=\pi/2$ and 
$u^{\theta}=0$, or
\begin{equation*}
-\left(1-\frac{2M}{r}\right)(u^{t})^{2}
-\frac{4aM}{r}u^{t}u^{\phi}
+\frac{r^{2}}{\Delta}(u^{r})^{2}
+\left(r^{2}+a^{2}+\frac{2Ma^{2}}{r}\right)(u^{\phi})^{2}=-1,
\end{equation*}
we obtain
\begin{equation}
\frac{1}{2}\dot{r}^{2}+V_{\rm eff}(r)=0,
\label{eq:eom}
\end{equation}
where the effective potential $V_{\rm eff}(r)$ is given by 
\begin{equation}
V_{\rm eff}(r)=-\frac{M}{r}+\frac{L^{2}-a^{2}(e^{2}-1)}{2r^{2}}-\frac{M(L-ae)^{2}}{r^{3}}-\frac{e^{2}-1}{2}.
\label{eq:effective_potential}
\end{equation}
The effective potential can be efficiently analyzed by 
introducing $y=M/r$. We put
\begin{eqnarray}
g(y)= -2(l-a_{*}e)^{2}y^{3}+[l^{2}-a_{*}^{2}(e^{2}-1)]y^{2}-2y-(e^{2}-1) 
\label{eq:gy}
\end{eqnarray}
and 
\begin{equation*}
D(y)=a_{*}^{2}y^{2}-2y+1.
\end{equation*}
Then, we have $V_{\rm eff}=g(y)/2$ and $\Delta=r^{2}D(y)$.
There are two positive roots of $D=0$ for $0< a_{*}\le 1$, which are
given by 
\begin{equation*}
y_{H}=\frac{1-\sqrt{1-a_{*}^{2}}}{a_{*}^{2}},\quad 
y_{C}=\frac{1+\sqrt{1-a_{*}^{2}}}{a_{*}^2},
\end{equation*} 
where $y_{H}=M/r_{H}$ and $y_{C}=M/r_{C}$ 
correspond to the event horizon and the Cauchy horizon, respectively.
These two roots coincide with each other at $y=1$ 
in the extremal case $a_{*}=1$. 
The region outside the horizon is transformed into $0<y<y_{H}$.
We should note the following useful relation
\begin{equation*}
\Omega_{H}=\frac{a_{*}y_{H}}{2M}.
\end{equation*}

For a particle which is initially at rest at infinity, i.e., 
marginally bound $e=1$, to reach the horizon,
the potential $g(y)$ must be nonpositive for $0<y<y_{H}$.
For $e=1$, the potential is given by 
\begin{equation*}
g(y)=-y[2(l-a_{*})^{2}y^{2}-l^{2}y+2].
\end{equation*}
Thus, the condition reduces to 
that $2(l-a_{*})^{2}y^{2}-l^{2}y+2$ is nonnegative.
After some straightforward 
calculation, we can obtain the following condition~\cite{Grib_Pavlov2010_Kerr}
\begin{equation*}
-2(1+\sqrt{1+a_{*}})=l_{L}\le l \le l_{R}=2(1+\sqrt{1-a_{*}}).
\end{equation*}
A similar condition also exists for the non-marginally bound case.
However, we should note that this does not apply if the particle 
scatters with other particles and changes its energy and 
angular momentum on the way to the horizon.

\subsection{CM energy of two particles in the Kerr spacetime}
We consider the collision of two particles 1 and 2 of 
the same rest mass $m_{0}$. 
We assume that the two particles are at the same spacetime point.
The four momentum of particle $i$  ($i=1,2$) is given by
\begin{equation*}
p_{i}^{a}=m_{0}u^{a}_{i},
\end{equation*}
where $u_{i}^{a}$ is the four velocity of particle $i$.
The sum of the two momenta is given by
\begin{equation*}
p_{\rm t}^{a}=p_{1}^{a}+p_{2}^{a}.
\end{equation*}
The CM energy $E_{\rm cm}$ of the two particles 
is then given by
\begin{equation}
E_{\rm cm}^{2}=-p^{a}_{\rm t}p_{{\rm t}a}=2m_{0}^{2}(1-g_{ab}u^{a}_{1}u_{2}^{b}).\label{eq:center-of-mass_energy}
\end{equation}

On the background metric (\ref{eq:equatorial_BLmetric}), 
using Eqs.~(\ref{eq:tdot}), (\ref{eq:phidot}), (\ref{eq:eom})
and (\ref{eq:gy}) in Eq.~(\ref{eq:center-of-mass_energy}),
the CM energy of two particles 1 and 2 
in the Kerr spacetime is calculated as
\begin{eqnarray}
\frac{E^{2}_{\rm cm}}{2m_{0}^{2}}
&=& 1-g_{tt}u^{t}_{1}u^{t}_{2}-g_{t\phi}(u^{t}_{1}u^{\phi}_{2}+u^{\phi}_{1}u^{t}_{2})-g_{rr}u^{r}_{1}u^{r}_{2}-g_{\phi\phi}u^{\phi}_{1}u^{\phi}_{2} \nonumber \\
&=&1-e_{1}e_{2} +\frac{
F(y)-G(y)}{D(y)},
\label{eq:ECOM_Kerr}
\end{eqnarray}
where $e_{i}$ and $l_{i}$ are $e$ and $l$ for particle $i$, 
\begin{eqnarray}
F(y)&=&2[a_{*}^{2}y^{2}(1+y)+(1-y)]e_{1}e_{2}
-2a_{*}y^{3}(e_{1}l_{2}+l_{1}e_{2})-(1-2y)y^{2}l_{1}l_{2}, 
\label{eq:Fy}\\
G&=&(\pm\sqrt{-g_{1}})(\pm\sqrt{-g_{2}}), 
\label{eq:Gy}\\
g_{i}(y)&=& -2(l_{i}-a_{*}e_{i})^{2}y^{3}+[l_{i}^{2}-a_{*}^{2}(e_{i}^{2}-1)]y^{2}-2y-
(e_{i}^{2}-1), 
\label{eq:giy} \\
D(y)&=&a_{*}^{2}y^{2}-2y+1,
\label{eq:Dy}
\end{eqnarray}
and the sign in front of $\sqrt{-g_{i}}$ 
in the expression of $G$ in Eq.~(\ref{eq:Gy}) 
corresponds to the sign of $u^{r}_{i}$.
In the following we assume $\dot{r}\le 0$ for
both particles and hence $G(y)=\sqrt{(-g_{1})(-g_{2})}$.

\section{Near-horizon collision around a Kerr black hole}

\subsection{Near-horizon collision around a subextremal Kerr black hole}
\label{subsec:subextreme}
We will see the near-horizon behavior of particles with the 
angular momentum $l=l_{H}$, which we call critical, 
and smaller angular momentum $l<l_{H}$, which we call subcritical.
We find 
\begin{equation}
g(y_{H})=-y_{H}^{2}(2e-a_{*}y_{H}l)^{2}=-a_{*}^{2}y_{H}^{4}(l_{H}-l)^{2},
\label{eq:gyH}
\end{equation}
noting that $l_{H}$ can be written in terms of $a_{*}$, $y_{H}$ and $e$ as
$l_{H}=2e/(a_{*}y_{H})$. Hence, $g(y_{H})\le 0$.
It is interesting to see whether a particle with $l=l_{H}$ 
which approaches the horizon is possible. With $l=l_{H}$, we 
have $g(y_{H})=0$, while 
\begin{equation}
g'(y_{H})=-2\frac{\sqrt{1-a_{*}^{2}}}{a_{*}^{2}}[(1+\sqrt{1-a_{*}^{2}})^{2}e^{2}+a_{*}^{2}],
\label{eq:gyHprime}
\end{equation}
where the prime denotes the differentiation with respect to the argument.
This is negative for the subextremal Kerr case $a_{*}^{2}<1$. This means that 
for the subextremal case, the effective potential $g(y)$ is positive 
in the vicinity of the horizon and hence 
a particle with the angular momentum $l=l_{H}$
is prohibited to approach the horizon.
On the other hand, 
there does exist a particle with slightly smaller angular 
momentum $l=l_{H}-\delta$ which approaches the horizon 
in the vicinity of the horizon. 
The CM energy for the collision involving this particle 
can be arbitrarily high in the limit $\delta\to 0$ even in the subextremal 
Kerr case~\cite{Grib_Pavlov2010_Kerr}.
For the subcritical orbit $l<l_{H}$, Eq.~(\ref{eq:gyH}) implies 
that $r$ is given near the horizon in terms of the 
particle's proper time $\tau$ as
\begin{equation}
r-r_{H}\simeq -a_{*}y_{H}^{2}(l_{H}-l)\tau+\mbox{const}.
\label{eq:proper_time_subcritical}
\end{equation} 
This means that for the fixed initial radius, a subcritical 
particle reaches the horizon after a proper time inversely 
proportional to $(l_{H}-l)$. 

Then, we will take the limit to the horizon in Eq.~(\ref{eq:ECOM_Kerr}) to 
consider the collision near the horizon. 
Noting 
\begin{eqnarray*}
F(y_{H})&=&a_{*}^{2}y_{H}^{4}(l_{H1}-l_{1})(l_{H2}-l_{2}), 
\end{eqnarray*}
where $l_{Hi}$ is the critical angular momentum $l_{H}$
for particle $i$  ($i=1,2$), 
combined with Eq.~(\ref{eq:gyH}), we can see that 
the terms of $O(1)$ in the numerator $F-G$ of the 
fraction on the right-hand side of 
Eq.~(\ref{eq:ECOM_Kerr}) cancel out. 
The nonvanishing contribution comes from the next order terms. Using l'Hospital's rule, the result is the following:
\begin{equation*}
\frac{E^{2}_{\rm cm}}{2m_{0}^{2}}
=1-e_{1}e_{2}+\lim_{y\to y_{H}}\frac{F'-G'}{D'}.
\end{equation*}
The derivatives are given by 
\begin{eqnarray*}
F'(y)&=&2[a_{*}^{2}(2y+3y^{2})-1]e_{1}e_{2}-6a_{*}y^{2}(e_{1}l_{2}+l_{1}e_{2})
-2y(1-3y)l_{1}l_{2}, \\
g_{i}'(y)&=& -6(l_{i}-a_{*}e_{i})^{2}y^{2}+2[l_{i}^{2}-a_{*}^{2}(e_{i}^{2}-1)]y-2, \\
D'(y)&=& 2(a_{*}^{2}y-1), \\
G'&=&G\frac{1}{2}\left(
\frac{g_{1}'}{g_{1}}+\frac{g_{2}'}{g_{2}}\right).
\end{eqnarray*}
From this form, we can see that there are two first-order 
poles, where $g_{i}(y_{H})=0$ for $i=1, 2$.
By implementing the calculation and taking the limit, 
we reach the following formula:
\begin{equation}
\frac{E_{\rm cm}}{2m_{0}}=
\sqrt{1+\frac{4[(l_{H1}-l_{1})-(l_{H2}-l_{2})]^{2}
+(l_{H1}l_{2}-l_{H2}l_{1})^{2}}{16(l_{H1}-l_{1})(l_{H2}-l_{2})}}.
\label{eq:general_subextremal}
\end{equation}
This is the formula for the CM energy of
two particles along the general geodesic orbits on the equatorial plane.
We should note that the right-hand side is given 
only in terms of the particles' angular momenta $l_{1}$ and $l_{2}$
and their critical values $l_{H1}$ and $l_{H2}$.
In terms of the quantities which have more 
direct physical meanings, Eq.~(\ref{eq:general_subextremal}) can be rewritten as follows:
\begin{equation}
\frac{E_{\rm cm}}{2m_{0}}=
\sqrt{1+\frac{4M^{2}[(e_{1}-\Omega_{H}L_{1})-(e_{2}-\Omega_{H}L_{2})]^{2}
+(e_{1}L_{2}-e_{2}L_{1})^{2}}{16M^{2}(e_{1}-\Omega_{H}L_{1})(e_{2}-\Omega_{H}L_{2})}}, 
\label{eq:general_subextremal_physical}
\end{equation}
where $L_{i}$ is $L$ for particle $i$.
In fact, as we will prove in Sec.~\ref{subsec:extreme}, Eq.~(\ref{eq:general_subextremal}) or equivalently Eq.~(\ref{eq:general_subextremal_physical}) 
is valid even for the extremal Kerr black hole simply by taking the 
near-extremal limit $a_{*}\to 1$. 
The necessary condition for obtaining an arbitrarily high $E_{\rm cm}$
is therefore $l\to l_{H}$ or $\Omega_{H}L\to e$
for either of the two 
particles. 

If we assume that only particle 1 is near-critical in 
Eq.~(\ref{eq:general_subextremal}), we obtain
\begin{equation}
\frac{E_{\rm cm}}{2m_{0}}\approx \sqrt{\frac{4+l_{H1}^{2}}{16}
\frac{l_{H2}-l_{2}}{l_{H1}-l_{1}}}.
\label{eq:single_critical}
\end{equation}

For $e_{1}=e_{2}=e$, we denote $l_{H1}=l_{H2}=l_{H}$ and
Eq.~(\ref{eq:general_subextremal}) reduces to 
\begin{equation}
\frac{E_{\rm cm}}{2m_{0}}=\sqrt{1+\frac{(l_{1}-l_{2})^{2}(4+l_{H}^{2})}{16(l_{H}-l_{1})(l_{H}-l_{2})}},
\label{eq:same_energy_subextremal}
\end{equation}
which reproduces the corresponding formula
in~\cite{Grib_Pavlov2010_Kerr}. 
When we set $e_{1}=e_{2}=1$, $l_{1}=l_{R}$ and $l_{2}=l_{L}$ in 
Eq.~(\ref{eq:same_energy_subextremal}), 
we obtain
\begin{equation}
\frac{E_{\rm cm}}{2m_{0}}
=\frac{1}{\sqrt[4]{1-a_{*}^{2}}}\sqrt{\frac{(1-a_{*}^{2})+(1+\sqrt{1+a_{*}}+\sqrt{1-a_{*}})^{2}}
{1+\sqrt{1-a_{*}^{2}}}}.
\label{eq:upper_bound_marginally_bound}
\end{equation}
This reproduces the corresponding formula in~\cite{Grib_Pavlov2010_Kerr}.
This provides an upper bound for the collision of two 
marginally bound particles.
For $a_{*}=0.998$, $E_{\rm cm}/(2m_{0})\simeq 9.49$ for this collision.
In the limit $a_{*}\to 1$, we have
\begin{equation}
\frac{E_{\rm cm}}{2m_{0}}\approx \frac{1+\sqrt{2}}{\sqrt[4]{1-a_{*}^{2}}}
\simeq \frac{2.41}{\sqrt[4]{1-a_{*}^{2}}},
\label{eq:bound_MB}
\end{equation}
which are given 
in~\cite{Berti_etal2009,Jacobson_Sotiriou2010,Grib_Pavlov2010_Kerr}.

If $e_{1}=1$ and $l_{1}=l_{R}$ for particle 1 and  
particle 2 takes a subcritical orbit, we obtain
\begin{equation}
\frac{E_{\rm cm}}{2m_{0}}\approx
\frac{1}{\sqrt{2}\sqrt{2-\sqrt{2}}}\frac{\sqrt{2e_{2}-l_{2}}}
{\sqrt[4]{1-a_{*}^{2}}}
\end{equation}
in the near-extremal limit.
Although the numerical factor depends on the choice of $e_{2}$ and 
$l_{2}$, 
the proportionality to $(1-a_{*}^{2})^{-1/4}$ does not change
as long as $e_{1}=1$ and $l_{1}=l_{R}$.

\subsection{Near-horizon collision around an extremal Kerr black hole}
\label{subsec:extreme}

For the extremal Kerr black hole $a_{*}^{2}=1$,
the effective potential is given by 
\begin{equation}
g(y)=-2(l-e)^{2}y^{3}+(l^{2}-e^{2}+1)y^{2}-2y-(e^{2}-1).
\label{eq:gyH_extremal}
\end{equation}
The double root $y=y_{H}=1$ of $D(y)$ gives an event horizon.
We should note that the region outside the horizon, $r>M$, is
transformed to $0<y<1$. 

From Eqs.~(\ref{eq:gyH}) and (\ref{eq:gyHprime}), for the 
critical orbit $l=l_{H}=2e$,  
we find $g(y_{H})=g'(y_{H})=0$ and hence $y=y_{H}$
is a stationary point of the effective potential. 
In fact, for the critical orbit, we have 
\begin{equation}
g(y)=-(3e^{2}-1)(1-y)^{2}+2e^{2}(1-y)^{3}.
\label{eq:critical_potential_extremal}
\end{equation}
Thus, there exists a critical orbit in the vicinity of the horizon if and 
only if $3e^{2}>1$ and then
the effective potential takes a maximum which is zero at $y=y_{H}=1$.
From this fact, one might infer an unstable circular orbit for 
a massive particle
at $y=y_{H}$, i.e., on the horizon which is a null hypersurface. 
This apparent paradox is resolved in Sec.~\ref{subsec:fake}. 
On the other hand, the maximal point on the horizon implies 
the existence of an orbit for a massive particle with $l=l_{H}$
which asymptotes the horizon. In fact, 
from Eq.~(\ref{eq:critical_potential_extremal}) we have 
\begin{equation}
\dot{r}=u^{r}=-\sqrt{3e^{2}-1}(1-y)\sqrt{1-\frac{2e^{2}}{3e^{2}-1}(1-y)}.
\label{eq:ur_critical_extremal}
\end{equation}
and this can be integrated to 
give
\begin{equation*}
\ln |r-M|\approx -\sqrt{3e^{2}-1}\frac{\tau}{M}+\mbox{const},
\end{equation*}
near the horizon. 
Thus, the critical particle approaches the horizon as $\tau\to \infty$,
as shown in~\cite{Jacobson_Sotiriou2010,Grib_Pavlov2010_Kerr}.
For the subcritical orbit $l<2e$, 
since 
\begin{equation*}
-g(y)=\left[(2e-l)-2(e-l)(1-y)\right]^{2}-\left[1-(e-l)(3e-l)\right](1-y)^{2}
-2(e-l)^{2}(1-y)^{3},
\end{equation*}
we have a different behavior of $u^{r}$ as
\begin{equation*}
\dot{r}=u^{r}=-\left[(2e-l)-2(e-l)(1-y)\right]
\sqrt{1-\frac{\left[1-(e-l)(3e-l)\right]+2(e-l)^{2}(1-y)}{\left[(2e-l)-2(e-l)(1-y)\right]^{2}}(1-y)^{2}}.
\end{equation*}
The proper time for the { subcritical} 
particle to reach the horizon is inversely proportional to $(2e-l)$
because Eq.~(\ref{eq:proper_time_subcritical}) is still valid even
in the extremal Kerr case.

If particle 1 takes a critical orbit but particle 2 takes a { subcritical}
orbit, 
the CM energy is given by the near-horizon limit of 
Eq.~(\ref{eq:ECOM_Kerr}) as
\begin{equation}
\frac{E_{\rm cm}}{2 m_{0}}\approx \sqrt{\frac{(2e_{2}-l_{2})
(2e_{1}-\sqrt{3e_{1}^{2}-1})}{2(1-y)}},
\label{eq:critical_extremal}
\end{equation}
where we have used Eq.~(\ref{eq:ur_critical_extremal}).
For the special case $e_{1}=e_{2}=e$, Eq.~(\ref{eq:critical_extremal}) 
reproduces the corresponding formula in~\cite{Grib_Pavlov2010_Kerr}.

If both particles take { subcritical} orbits, 
the fraction on the right-hand side of Eq.~(\ref{eq:ECOM_Kerr}) is 
bounded because both the numerator $F-G$ and the denominator $D$
have a second-order zero at the horizon { $y=y_{H}=1$}.
Estimating the terms of $O((1-y)^{2})$ in $F-G$ and $D$
by Taylor series expansion, we obtain
\begin{equation}
\frac{E_{\rm cm}}{2m_{0}}=\sqrt{
\frac{1}{2}\left[1-e_{1}e_{2}+\frac{2e_{2}-l_{2}}{2e_{1}-l_{1}}\frac{1+e_{1}^{2}}{2}
+\frac{2e_{1}-l_{1}}{2e_{2}-l_{2}}\frac{1+e_{2}^{2}}{2}\right]}.
\label{eq:general_extremal}
\end{equation}
More systematically, we can take the following approach.
For the extremal case, the numerator $F-G$ and the denominator $D$ 
both must have a second-order zero at $y=y_{H}$. Using l'Hospital's 
rule twice, we obtain
\begin{equation}
\frac{E^{2}_{\rm cm}}{2m_{0}^{2}}
=1-e_{1}e_{2}+\lim_{y\to y_{H}}\frac{F''-G''}{D''},
\end{equation}
where the second-order derivatives are given by 
\begin{eqnarray*}
F''(y)&=&4[a_{*}^{2}(1+3y)]e_{1}e_{2}-12a_{*}y(e_{1}l_{2}+l_{1}e_{2})
-2(1-6y)l_{1}l_{2}, \\
g''_{i}(y)&=& -12(l_{i}-a_{*}e_{i})^{2}y+2[l_{i}^{2}-a_{*}^{2}(e_{i}^{2}-1)], \\
D''(y)&=& 2a_{*}^{2}, \\
G''&=&G\left[\frac{1}{2}\left(
\frac{g_{1}''}{g_{1}}+\frac{g_{2}''}{g_{2}}\right)
-\frac{1}{4}\left(\frac{g_{1}'}{g_{1}}-\frac{g_{2}'}{g_{2}}\right)^{2}
\right]
\end{eqnarray*}
and $a_{*}=1$ for the extremal case.
It is found that this approach also yields Eq.~(\ref{eq:general_extremal}).

In the course of derivation, it is not so obvious
whether the formula for the subextremal Kerr black hole
given by Eq.~(\ref{eq:general_subextremal}) reproduces 
the formula (\ref{eq:general_extremal})
for the extremal case if we take the near-extremal limit $a_{*}\to 1$
in the former.
In fact, it is not difficult to see that this is the case by putting 
$l_{Hi}=2e_{i}$ in Eq.~(\ref{eq:general_subextremal}).
Therefore, the general formula (\ref{eq:general_subextremal}) or
(\ref{eq:general_subextremal_physical}), which has been derived 
for the subextremal case, is applicable in both the 
subextremal and extremal cases.  

We can confirm that for the special case $e_{1}=e_{2}=e$,
Eq.~(\ref{eq:general_extremal}) reduces to
\begin{equation*}
\frac{E_{\rm cm}}{2m_{0}}=\sqrt{1+\frac{1+e^{2}}{4}
\frac{(l_{1}-l_{2})^{2}}{(2e-l_{1})(2e-l_{2})}},
\end{equation*}
which coincides with the corresponding formula 
in~\cite{Grib_Pavlov2010_Kerr}.  
Moreover, Eq.~(\ref{eq:general_extremal}) reduces to
\begin{equation*}
\frac{E_{\rm cm}}{2m_{0}}=\sqrt{\frac{1}{2}\left(
\frac{2-l_{1}}{2-l_{2}}
+\frac{2-l_{2}}{2-l_{1}}\right)},
\end{equation*}
for the special case $e_{1}=e_{2}=1$,
which reproduces the formula discovered by BSW~\cite{BSW2009}.

\subsection{The circular timelike orbit on the 
extremal Kerr black hole horizon is fake}
\label{subsec:fake}

As we have seen in Sec.~\ref{subsec:extreme},
in the extremal Kerr case, 
$r=r_{H}$ is a zero and maximal point of the 
effective potential.
{ From this fact, one might infer that 
an unstable circular orbit for a massive particle 
is possible at $r=r_{H}$.}  
However, we will show that this is not real.

We should note that the 
Boyer-Lindquist coordinate system has a coordinate singularity at 
$r=r_{H}$. To avoid the complication due to the coordinate singularity,
we move to the ingoing Kerr coordinates~\cite{Poisson2004}:
\begin{equation*}
dv=dt+(r^{2}+a^{2})\frac{dr}{\Delta}, \quad 
d\varphi =d\phi+a\frac{dr}{\Delta}.
\end{equation*}
The line element then can be written as
\begin{eqnarray*}
ds^{2}&=&-\left(1-\frac{2Mr}{\rho^{2}}\right) dv^{2}+2dvdr+\rho^{2}d\theta^{2}
+\frac{[(r^{2}+a^{2})^{2}-a^{2}\Delta \sin^{2}\theta]\sin^{2}\theta}{\rho^{2}}d\varphi ^{2} \nonumber \\
&& -2a \sin^{2}\theta d\varphi dr-\frac{4aMr}{\rho^{2}}\sin^{2}
\theta d\varphi dv.
\end{eqnarray*}
The Killing vectors are given by
\begin{equation*}
\xi^{a}=\left(\frac{\partial}{\partial v}\right)^{a},\quad 
\psi^{a}=\left(\frac{\partial}{\partial \varphi}\right)^{a}.
\end{equation*}
On the equatorial plane $\theta=\pi/2$, the line element 
in the extremal Kerr { spacetime} is given by
\begin{equation*}
ds^{2}=-\left(1-\frac{2M}{r}\right)dv^{2}+2dvdr+
\left(r^{2}+M^{2}+\frac{2M^{3}}{r}\right)d\varphi ^{2}-2Md\varphi dr
-\frac{4M^{2}}{r}d\varphi dv.
\end{equation*}
The conserved quantities are given by 
\begin{eqnarray}
e&=&-g_{ab}\xi^{a}u^{b}=\left(1-\frac{2M}{r}\right)\dot{v}-\dot{r}+\frac{2M^{2}}{r}\dot{\varphi }, 
\label{eq:e_ingoing_Kerr}\\
L&=&g_{ab}\psi^{a}u^{b}=\left(r^{2}+M^{2}+\frac{2M^{3}}{r}\right)\dot{\varphi }-M\dot{r}-\frac{2M^{2}}{r}\dot{v}.
\label{eq:l_ingoing_Kerr}
\end{eqnarray}

Putting $r=r_{H}=M$ in Eqs.~(\ref{eq:e_ingoing_Kerr}) 
and (\ref{eq:l_ingoing_Kerr}), we have
\begin{eqnarray}
e&=&-\dot{v}-\dot{r}+2M\dot{\varphi }, 
\label{eq:energy_horizon}\\
L&=& -2M\dot{v}-M\dot{r}+4M^{2}\dot{\varphi }.
\label{eq:angular_momentum_horizon}
\end{eqnarray}
The norm of $u^{a}$ can be written at $r=M$ as
\begin{equation}
u^{a}u_{a}=\dot{v}^{2}+2\dot{r}\dot{v}+4M^{2}\dot{\varphi }^{2}-2M\dot{\varphi }
\dot{r}-4M\dot{\varphi }\dot{v}.
\label{eq:norm}
\end{equation}
If we assume that the particle remains on the horizon $r=M$, we have $\dot{r}=0$
and we may conclude $L/e=2M$ or $e=L=0$ from 
Eqs.~(\ref{eq:energy_horizon}) and (\ref{eq:angular_momentum_horizon}).
On the other hand, { one} cannot 
solve Eqs.~(\ref{eq:energy_horizon}) and 
(\ref{eq:angular_momentum_horizon}) 
for $\dot{v}$ and $\dot{\varphi }$ separately 
in terms of $e$ and $L$ 
because of the degeneracy and hence cannot obtain Eqs.~(\ref{eq:eom})
and (\ref{eq:effective_potential}).
In other words, Eqs.~(\ref{eq:eom}) and (\ref{eq:effective_potential}) 
do not make sense in the present case.
Instead, we obtain from 
Eqs.~(\ref{eq:energy_horizon}) and (\ref{eq:norm})
\begin{equation*}
u^{a}u_{a}=e^{2}.
\end{equation*}
{ Since $u^{a}u_{a}\le 0$ for causal geodesics, 
this means that a causal geodesic can remain on the horizon only if 
it is a null geodesic with $e=L=0$.
Thus, the timelike circular orbit on the horizon, 
which might be inferred from the 
stationary point of the effective potential at $r=r_{H}$, is fake.
Note also that the angular velocity of this null geodesic 
which remains on the horizon is given by 
\begin{equation*}
\frac{d\varphi }{dv}=\frac{\dot{\varphi }}{\dot{v}}=\frac{1}{2M}=\Omega_{H},
\end{equation*}
indicating that this null geodesic is a generator of the event horizon.}

This is of course entirely consistent with the following general argument.
Any timelike curve cannot remain on an event horizon because 
the event horizon is normal to the Killing vector $\chi^{a}$, which is 
null on the horizon, and hence the tangent space at a point on the horizon is 
spanned by $\chi^{a}$ and two spacelike vectors $e_{(1)}^{a}$
and $e_{(2)}^{a}$ which are orthogonal to $\chi^{a}$. 
Any linear combination of the null vector $\chi^{a}$ and the 
spacelike vectors $e_{(1)}^{a}$ and $e_{(2)}$ is either null or spacelike. 

\section{Near-horizon collision of a particle plunging from the ISCO}
As is emphasized in~\cite{Grib_Pavlov2010_Kerr} and in Sec.~\ref{subsec:particle_orbits}, the upper bound~(\ref{eq:bound_MB}) applies only for the particles 
which begin at rest at infinity and 
reach the horizon all the way from infinity along the geodesic.
If a particle loses or gains its energy or angular momentum on 
the way to the horizon, this limit does not apply.
In this respect, particles plunging from the ISCO are 
considered very natural as particles plunging into 
the horizon in an astrophysical context.

The ISCO in the Kerr spacetime is explicitly given by Bardeen, Press 
and Teukolsky~\cite{BPT1972}.
The circular orbit on the equatorial plane in the Kerr metric 
is given by 
$V_{\rm eff}(r)=V_{\rm eff}'(r)=0$, where the prime denotes
the derivative with respect to the argument.
The condition implies 
\begin{eqnarray}
e&=&\frac{r^{1/2}(r-2M)+sa M^{1/2}}
{r^{3/4}(r^{3/2}-3M r^{1/2}+s2aM^{1/2})^{1/2}}, 
\label{eq:circular_orbit_energy}
\\
L&=&s\frac{M^{1/2}(r^{2}+a^{2}-s2M^{1/2}ar^{1/2})}
{r^{3/4}(r^{3/2}-3M r^{1/2}+s2aM^{1/2})^{1/2}},
\label{eq:circular_orbit_angular_momentum}
\end{eqnarray}
where we have assumed $0\le a< M$ and $s=1$ and $-1$
correspond to the prograde and retrograde orbits, respectively.
The ISCO is determined by the 
condition $de/dr=dL/dr=0$. The radius of the ISCO is then given 
by
\begin{eqnarray}
\frac{r_{\rm ISCO}}{M}&=&
3+Z_{2}-s[(3-Z_{1})(3+Z_{1}+2Z_{2})]^{1/2}, 
\label{eq:r_isco}\\
Z_{1}&=&1+(1-a_{*}^{2})^{1/3}[(1-a_{*})^{1/3}+(1+a_{*})^{1/3}], \quad 
Z_{2}=(3a_{*}^{2}+Z_{1}^{2})^{1/2}, 
\label{eq:Z1_Z2}
\end{eqnarray}
where $0\le a_{*}< 1$. 
The energy and angular momentum of the particle at the ISCO are 
calculated by substituting Eqs.~(\ref{eq:r_isco})
and (\ref{eq:Z1_Z2})
for $r=r_{\rm ISCO}$ into 
Eqs.~(\ref{eq:circular_orbit_energy}) and 
(\ref{eq:circular_orbit_angular_momentum}).

To see the behavior in the near-extremal limit $a_{*}\to 1$, 
we put $a_{*}=1-\epsilon$ and expand the above obtained expression 
in terms of $\epsilon$. From Eqs.~(\ref{eq:Z1_Z2}), we obtain
\begin{eqnarray*}
Z_{1}= 1+2^{2/3}\epsilon^{1/3}+2^{1/3}\epsilon^{2/3}+O(\epsilon), \quad 
Z_{2}= 2+\frac{1}{2}2^{2/3}\epsilon^{1/3}+\frac{7}{8}2^{1/3}\epsilon^{2/3}
+O(\epsilon).
\end{eqnarray*}
Then, using Eqs.~(\ref{eq:circular_orbit_energy}), (\ref{eq:circular_orbit_angular_momentum}) and (\ref{eq:r_isco}), we obtain 
\begin{eqnarray}
\frac{r_{\rm ISCO}}{M}&=&1+2^{2/3}\epsilon^{1/3}+\frac{7}{4}2^{1/3}\epsilon^{2/3}+O(\epsilon), 
\label{eq:r_isco_near-extreme}\\
e&=&\frac{\sqrt{3}}{3}+\frac{2^{2/3}}{3}\sqrt{3}\epsilon^{1/3}-\frac{5}{12}2^{1/3}\sqrt{3}\epsilon^{2/3}+O(\epsilon) , 
\label{eq:e_isco_near-extreme}\\
l&=&\frac{2}{3}\sqrt{3}+\frac{2}{3}2^{2/3}\sqrt{3}\epsilon^{1/3}+\frac{1}{6}2^{1/3}\sqrt{3}\epsilon^{2/3}+O(\epsilon), 
\label{eq:l_isco_near-extreme}
\end{eqnarray}
and hence
\begin{equation*}
\frac{l}{e}= 2+3\cdot 2^{1/3}\epsilon^{2/3}+O(\epsilon)
\end{equation*}
for the prograde orbit ($s=1$).
For the retrograde orbit ($s=-1$), we obtain
\begin{eqnarray*}
\frac{r_{\rm ISCO}}{M}&=&9+O(\epsilon), \\
e&=&\frac{5}{9}\sqrt{3}+O(\epsilon), \\
l&=&-\frac{22}{9}\sqrt{3}+O(\epsilon), 
\end{eqnarray*}
and hence
\begin{equation*}
\frac{l}{e}=-\frac{22}{5}+O(\epsilon).
\end{equation*} 
On the other hand, $l_{H}$ can be 
written as 
\begin{equation*}
\frac{l_{H}}{e}=\frac{2a_{*}}{1-\sqrt{1-a_{*}^{2}}}=2+2\sqrt{2}\epsilon^{1/2}
+O(\epsilon).
\end{equation*}
Therefore, the prograde ISCO
particle has the angular momentum which coincides with the 
critical value $l_{H}$
in the near-extremal limit.

We should note that for the marginally bound particle 
with $l=l_{R}$ we have $l/e=l_{R}=2+2\epsilon^{1/2}$.  
Therefore, $(l_{H}-l)/e \approx 2\sqrt{2}\epsilon^{1/2}$ for the 
prograde ISCO, while $(l_{H}-l)/e \approx 2(\sqrt{2}-1)\epsilon^{1/2}$ 
for a marginally bound particle with $l=l_{R}$.
Since all other factors are nonzero finite, 
a particle which plunges from the prograde ISCO collides with 
a generic particle with $E_{\rm cm}\propto \epsilon^{-1/4}$,
as a marginally bound particle with $l=l_{R}$ does.
It also follows that if a particle plunging from the prograde ISCO
collides with a marginally bound particle with $l=l_{R}$, 
the CM energy is bounded even in 
the near-extremal limit $a_{*}\to 1$.

Using Eq.~(\ref{eq:single_critical}) with Eqs.~(\ref{eq:r_isco_near-extreme})--(\ref{eq:l_isco_near-extreme}), we can easily 
estimate the CM energy { near the horizon} 
for the near-extremal Kerr black hole.
If particle 1 is a particle plunging from the prograde ISCO and 
particle 2 takes a { subcritical} orbit, we obtain
\begin{equation}
\frac{E_{\rm cm}}{2m_{0}}\approx \frac{1}{2^{1/2}3^{1/4}}
\frac{\sqrt{2e_{2}-l_{2}}}{\sqrt[4]{1-a_{*}^{2}}}.
\end{equation}
Thus, the CM energy can be unboundedly high in the limit $a_{*}\to 1$.
Since the dependence $(1-a_{*})^{-1/4}$ is common to 
the upper bound (\ref{eq:bound_MB}) for marginally bound particles, 
we can conclude that the BSW 
effect occurs for a particle plunging from the prograde ISCO and
in this case the fine-tuning of the energy and the angular momentum 
is naturally realized in 
{ the standard accretion disks} with electromagnetic radiation 
or in inspiralling binaries with gravitational wave radiation. 
{ The ratio of $E_{\rm cm}$ for the ISCO particle to that for the 
marginally bound particle with $l=l_{R}$ is given by $\sqrt{2-\sqrt{2}}/3^{1/4}\simeq 0.582$ in the near-extremal limit.}

In the following, we consider 
the near-horizon collisions of a particle plunging 
from the prograde ISCO with (a) 
a marginally bound particle with $l=l_{L}$, 
(b) a particle plunging from the retrograde ISCO, 
and (c) a marginally bound particle with $l=0$.
In the near-extremal limit $a_{*}\to 1$, we obtain 
\begin{equation*}
\frac{E_{\rm cm}}{2m_{0}}\approx \frac{\alpha}{\sqrt[4]{1-a_{*}^{2}}}
\end{equation*}
where the numerical factor $\alpha=\sqrt{2e_{2}-l_{2}}/(2^{1/2}3^{1/4})$
is calculated to be $\sqrt{(2+\sqrt{2})/\sqrt{3}}\simeq 1.40$, 
$4/3\simeq 1.33$ and $1/\sqrt[4]{3}\simeq 0.760$
for cases (a), (b), and (c), respectively.

\begin{figure}[htbp]
\includegraphics[width=0.8\textwidth]{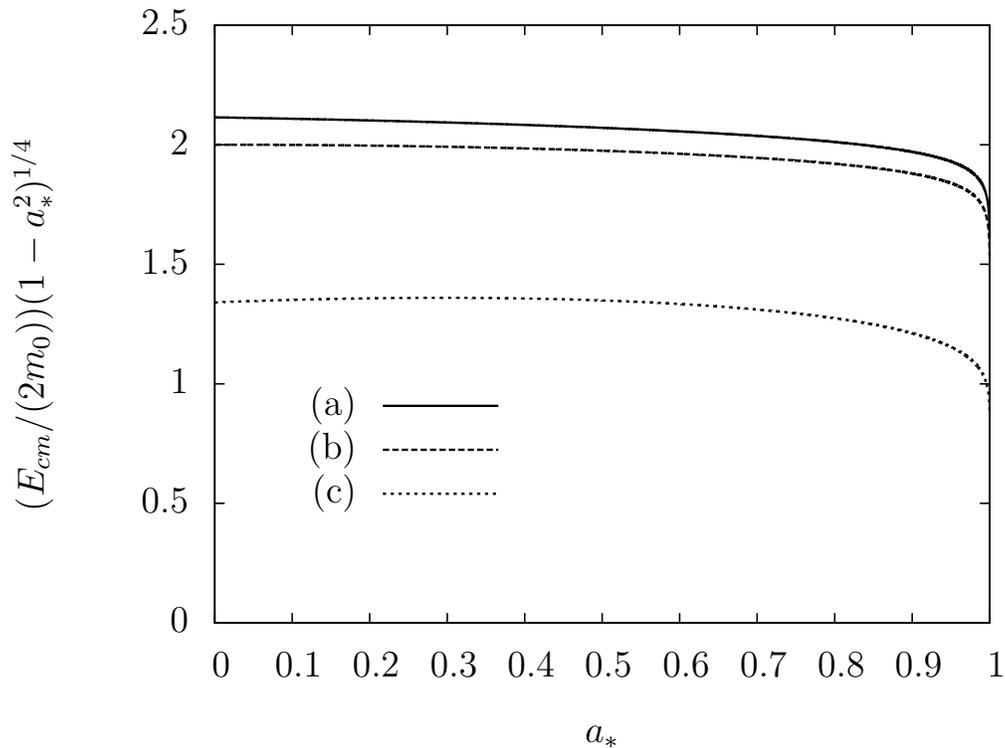}
\caption{\label{fg:collision_energy_maker}
The CM energy $E_{\rm cm}$ for the 
near-horizon collision.
The solid, dashed, and dotted curves denote
the collisions of a particle plunging from the prograde ISCO with
(a) a marginally bound particle with $l=l_{L}$, 
(b) a particle plunging from the retrograde ISCO, 
and (c) a marginally bound particle with $l=0$, respectively. 
For clarity, $E_{\rm cm}$ is multiplied by $\sqrt[4]{1-a_{*}^{2}}$ 
in the vertical axis.
}
\end{figure}

For the general values of $a_{*}$ in $0\le a_{*}<1$, we can calculate the 
CM energy using Eq.~(\ref{eq:general_subextremal}) with 
Eqs.~(\ref{eq:circular_orbit_energy})--(\ref{eq:Z1_Z2}). 
The result is summarized 
in Fig.~\ref{fg:collision_energy_maker}, where 
$E_{\rm cm}/(2m_{0})$ is multiplied by $\sqrt[4]{1-a_{*}^{2}}$
for clarity. 
In this figure, the solid, dashed, and dotted curves denote
cases (a), (b), and (c), respectively.
The  
CM energy for two particles, either of which is 
a particle plunging from the prograde ISCO
is always below the upper bound (\ref{eq:upper_bound_marginally_bound})
for marginally bound particles.
We should note that the dependence of 
$(E_{\rm cm}/(2m_{0})) (1-a_{*}^{2})^{1/4}$  on $a_{*}$ in 
$0\le a_{*}<1$ is very weak for cases (a), (b), and (c) and 
hence the empirical formula 
$E_{\rm cm}/(2m_{0})\sim 
1/\sqrt[4]{1-a_{*}^{2}}$ is a 
very good approximation within a factor of 2 or so
for all values of $a_{*}$ in $0\le a_{*}<1$.
Thus, this formula provides the typical value for the 
CM energy for the near-horizon collision of a particle 
which plunges from the prograde ISCO with a generic { subcritical} particle.
For the near-maximal rotation, the maximum value for 
the CM energy coincides with
the upper bound for a marginally bound particle within a factor of 2.

If we use Thorne's bound $a_{*}=0.998$ for the spin parameter, 
$E_{\rm cm}/(2m_{0})$
is calculated to be 6.95, 6.61, and 3.86 for cases (a), (b), and (c), 
respectively. 
This means that a highly relativistic collision 
can naturally occur near the horizon of a rapidly rotating black hole
in an astrophysical context.
{ Note that with highly relativistic collision we here 
mean systems where the CM energy is much larger than the rest mass.}

\section{Particle collision of a particle orbiting the ISCO}
 
In this section, we { deviate somewhat}
from the original idea of BSW~\cite{BSW2009}.
We consider the situation where 
a particle orbiting the ISCO collides with another particle
on the ISCO instead of a near-horizon collision.
In this case, we cannot take the near-horizon limit beforehand.
Although we do not expect a compact expression for the general case, 
we can obtain a simple formula for the near-extremal limit.
Using Eq.~(\ref{eq:ECOM_Kerr}) with 
Eqs.~(\ref{eq:Fy})--(\ref{eq:Dy})
and (\ref{eq:r_isco_near-extreme})--(\ref{eq:l_isco_near-extreme}) and 
$y=M/r_{\rm ISCO}$, we can 
estimate the CM energy for the near-extremal Kerr black hole.
If particle 1 is a particle orbiting the prograde ISCO and 
particle 2 takes a { subcritical} orbit, we obtain
\begin{equation}
\frac{E_{\rm cm}}{2m_{0}}\approx \frac{1}{2^{1/6}3^{1/4}}
\frac{\sqrt{2e_{2}-l_{2}}}{\sqrt[6]{1-a_{*}^{2}}}.
\end{equation}
It is quite intriguing that the dependence $(1-a_{*}^{2})^{-1/6}$ 
on the spin parameter { here} is quite
different from that for the near-horizon collision $(1-a_{*}^{2})^{-1/4}$. 
The CM energy can be arbitrarily high
in the near-extremal limit $a_{*}\to 1$.
However, this needs to be distinguished from the BSW 
effect for the near-horizon collision of plunging particles.
 
As in the near-horizon case, we consider 
the { ``on-ISCO''} collisions of a particle orbiting 
the prograde ISCO with (a) 
a marginally bound particle with $l=l_{L}$, 
(b) a particle plunging from the retrograde ISCO,
and (c) a marginally bound particle with $l=0$.
In the near-extremal limit $a_{*}\to 1$, we obtain 
\begin{equation*}
\frac{E_{\rm cm}}{2m_{0}}\approx \frac{\beta}{\sqrt[6]{1-a_{*}^{2}}}
\end{equation*}
where the numerical factor $\beta=\sqrt{2e_{2}-l_{2}}/(2^{1/6}3^{1/4})$
is calculated to be $\sqrt{2(2+\sqrt{2})}/(2^{1/6}3^{1/4})\simeq 1.77$, 
$4\cdot 2^{1/3}/3\simeq 1.68$, and 
$\sqrt{2}/(2^{1/6}3^{1/4})\simeq 0.957$
for cases (a), (b), and (c), respectively.

\begin{figure}[htbp]
\includegraphics[width=0.8\textwidth]{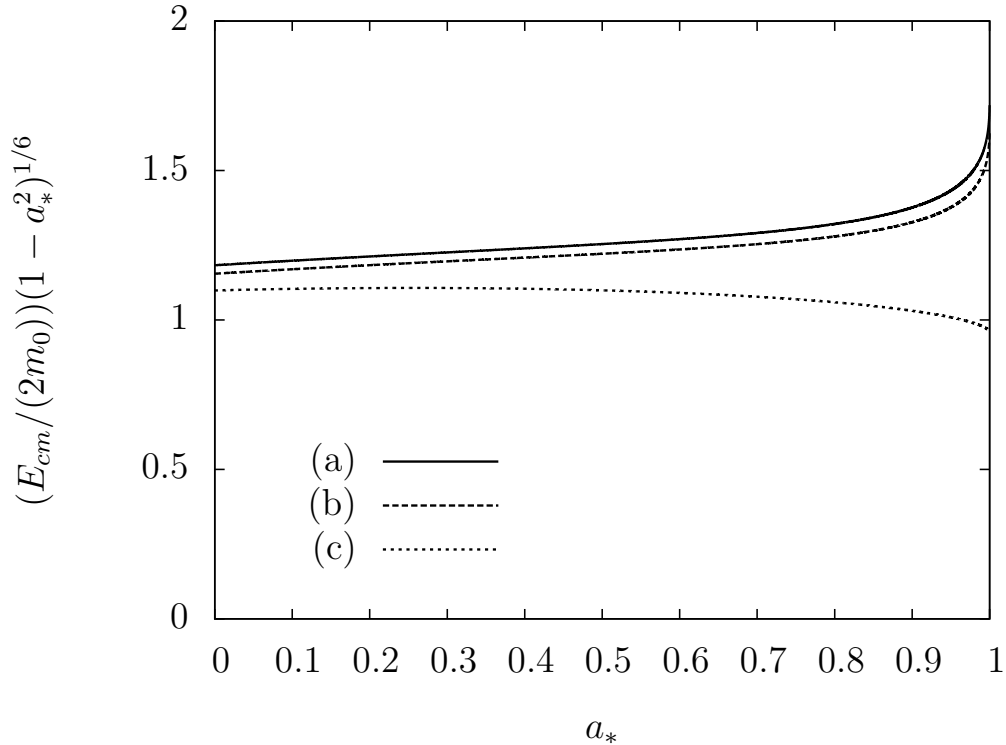}
\caption{\label{fg:collision_isco_maker}
The CM energy $E_{\rm cm}$ for the on-ISCO collision.
The solid, dashed, and dotted curves denote
the collisions of a particle orbiting the prograde ISCO with  
(a) a marginally bound particle with $l=l_{L}$, 
(b) a particle plunging from the retrograde ISCO, 
and (c) a marginally bound particle with $l=0$, respectively. 
For clarity, $E_{\rm cm}$ is multiplied by $\sqrt[6]{1-a_{*}^{2}}$ 
in the vertical axis.}
\end{figure}

For the general values of $a_{*}$ in $0\le a_{*}<1$, we can calculate the 
CM energy using Eq.~(\ref{eq:ECOM_Kerr}) with Eqs.~(\ref{eq:Fy})--(\ref{eq:Dy}) and (\ref{eq:circular_orbit_energy})--(\ref{eq:Z1_Z2}), $r=r_{\rm ISCO}$ and $y=M/r_{\rm ISCO}$. 
The result is summarized 
in Fig.~\ref{fg:collision_isco_maker}, where 
$E_{\rm cm}/(2m_{0})$ is multiplied by $\sqrt[6]{1-a_{*}^{2}}$
for clarity. In this figure, the solid, dashed, and dotted curves denote
cases (a), (b), and (c), 
respectively. 
The CM energy for two particles, either of which is 
a particle orbiting the prograde ISCO, 
is always below the upper bound (\ref{eq:upper_bound_marginally_bound})
for marginally bound particles.
We should note that the dependence of 
$(E_{\rm cm}/(2m_{0})) (1-a_{*}^{2})^{1/6}$  on $a_{*}$ in 
$0\le a_{*}<1$ is very weak and hence the empirical formula
$E_{\rm cm}/(2m_{0})\sim 
1/\sqrt[6]{1-a_{*}^{2}}$ is a 
very good approximation within a factor of 2 
for all values of $a_{*}$ in $0\le a_{*}<1$.
Thus, this formula provides the typical value for the 
CM energy for the { on-ISCO collision} of a particle 
which orbits the prograde ISCO with a generic { subcritical} particle.
It should be noted that the CM energy for the 
{ on-ISCO collision} is always smaller than that 
for the near-horizon collision of a particle plunging from 
the ISCO in the near-extremal limit
because of the different dependence on $a_{*}$.

If we use Thorne's bound $a_{*}=0.998$ for the spin parameter, 
$E_{\rm cm}/(2m_{0})$
is calculated to be 4.11, 3.91, and 2.43 for cases (a), (b), and (c), 
respectively. 
This means that a highly or moderately relativistic collision 
can naturally occur on the ISCO { around} a rapidly rotating black hole
in an astrophysical context.

\section{conclusion and discussion}
We have investigated particle collisions near the horizon 
and on the ISCO { around} a Kerr
black hole. We have derived a general explicit 
formula for the CM energy near the horizon 
in terms of the energies and the 
angular momenta of colliding two particles 
on the equatorial plane. 
We have confirmed that the obtained formula includes 
known formulas 
as its special cases.
We have explicitly shown that 
{ although the effective potential around a
maximally rotating Kerr black hole has a zero and maximal point 
on the horizon, it}
does not correspond to a real circular 
orbit. Then, we have studied the near-horizon collision of 
particles, either of which plunges from the ISCO.
We have shown that the BSW effect occurs for such a collision 
in the near-maximal rotation limit and that 
the maximum value for the CM energy
is the same within a factor of 2 as the upper bound 
for the marginally bound  
particles for which the angular momentum must be fine-tuned. 
We have also investigated the collision of a particle orbiting 
the ISCO with another generic particle on the ISCO 
and found that it is also the case
that one can obtain an arbitrarily high CM energy 
in the near-maximal rotation limit, 
although this energy is smaller than 
the value for the near-horizon collision in { this limit}.
The result implies that the BSW effect, which was originally proposed 
for the marginally bound particles with the fine-tuned angular momentum,
is astrophysically relevant since the fine-tuning is 
naturally realized for ISCO particles 
in { the standard accretion disks} and extreme mass-ratio inspirals.
Although the CM energy is bounded if the spin parameter 
of the black hole is bounded in an astrophysical context, 
the collision can still be highly or moderately relativistic near 
the horizon and on the ISCO { around} a rapidly rotating black hole.

The present result naively suggests the following scenario.
A highly or moderately
relativistic collision often occurs near the 
horizon of a rapidly rotating 
black hole in the context of the accretion disks and the extreme
mass-ratio binaries.
For the standard accretion disk, 
gamma rays with energy of several GeVs can be produced 
inside and around the inner edge of the disk, if the CM energies of 
protons and ions collisions are eventually converted to photons. 
These photons can have much higher energy than usual thermal
photons.

What is more intriguing is the high-velocity collision 
of compact objects around a 
supermassive or intermediate-mass black hole. 
Here, the compact objects will collide near the horizon or on the 
ISCO with a high ``relativistic gamma factor'' $E_{\rm cm}/(2m_{0})$.
The result will strongly depend on the kinds of the compact objects 
and the value of the relativistic gamma factor.  
For example, if two neutron stars collide with a sufficiently high gamma, it will result in the gravitational collapse to a black hole (e.g.~\cite{Miller_etal2001}). If two white dwarfs collide with a sufficiently high gamma,  they might be smashed, destroyed, and scattered away because of the CM energy much 
greater than the binding energy of the white dwarfs.
Thus, the collision of the compact objects around a rapidly 
rotating supermassive or intermediate-mass black hole
provides a unique laboratory for the relativistic collision 
of black holes, neutron stars, and white dwarfs.
The interaction of the compact object with the fluid or plasma 
near the horizon or on the ISCO might {also be} a striking phenomenon.
The details of all these processes would not be so simple and should be 
investigated not only by analytical arguments but also 
by numerical simulations, including numerical relativity and 
general relativistic 
hydrodynamics.

Finally, we speculate that peculiar signals 
originating from the highly or moderately relativistic collision
of particles, fluids and compact objects 
around a rapidly rotating black hole
might be detected 
by the direct observation of black holes 
by means of {electromagnetic 
and/or gravitational waves and/or neutrinos.}
For example, if two black holes collide with a sufficiently high gamma, a considerable fraction (as large as $14\pm 3$\% for head-on collision~\cite{Sperhake_etal2008} and $35\pm 5$\% for zoom-whirl 
collision~\cite{Shibata_etal2008,Sperhake_etal2009})
of the CM energy can be 
radiated away through gravitational radiation.
Of course, because of the strong redshift,
we cannot immediately expect that the energetic 
radiation can directly reach us.
However, the emission peculiar to such relativistic collisions
will be redshifted and might still be observed 
{in electromagnetic and/or gravitational waves 
and/or neutrinos}. 
In this respect, the {on-ISCO collision} might be 
more advantageous to observation than the near-horizon collision.
To investigate what signals would be observed from 
a highly or moderately relativistic 
collision, numerical simulations will be very powerful.

\acknowledgments
The authors {thank} U.~Miyamoto, K.~Nakao, R.~Takahashi, K.~Hioki, 
M.~Saijo, and M.~Shibata for fruitful discussion. 
The authors also thank the anonymous referee for helpful comments.
T.H. was supported by a Grant-in-Aid for Scientific Research from the Ministry of Education, Culture, Sports, Science and Technology {of} Japan [Young Scientists (B) No. 21740190].

\end{document}